\documentclass[
 reprint,
nofootinbib,
 amsmath,amssymb,
 aps,
]{revtex4-2}

\usepackage{graphicx}
\usepackage{dcolumn}
\usepackage{bm}
\usepackage{amsmath}
\usepackage{feynmp-auto}
\usepackage{braket,mleftright}
\usepackage{hyperref}
\usepackage{subcaption}
\begin{document}

\title{Determining the Meson Cloud Contribution of Nucleon \\Electromagnetic Form Factor Using Dispersion Relation}

\author{Ek-ong Atthaphan}
\email[]{ekorattha@gmail.com}
\author{Attaphon Kaewsnod}
\author{Kai Xu}
\author{Moh Moh Aung}
\author{Warintorn Sreethawong}
\author{Ayut Limphirat}
\email[]{ayut@sut.ac.th}
\author{Yupeng Yan}%
\email[]{yupeng@sut.ac.th}
\affiliation{%
 School of Physics and Center of Excellence in High Energy Physics and Astrophysics, Suranaree University of Technology, Nakhon Ratchasima 30000, Thailand }%

\date{\today}

\begin{abstract}
We explore the meson cloud contribution to nucleon electromagnetic form factors in dispersion relation approach. In our calculations, experimental data on transition amplitudes for pion-nucleon scatterings are taken directly as inputs, with the assumption that the photon-pion interaction dominates over other meson-photon couplings. Combining with the quark core contribution evaluated in quark model, the proton electromagnetic form factors are well reproduced in the work. The resulting mean-square charge radius of the proton agrees well with experimental data, and the quark core radius is reasonable.

\end{abstract}

\maketitle


\section{\label{sec:1intro}Introduction}

Understanding the internal structure of nucleons is a fundamental goal in nuclear and particle  physics. The bag model is among the most successful pictures, where the constituent quarks are sealed in a bag surrounded by a meson cloud \cite{BROWN200285_mc1,PhysRevD.22.2838_mc3,Thomas:2001kw_mc6}. The meson cloud plays an important role in low-energy scatterings, where the electromagnetic form factors are sensitive to the nucleon properties like charge, magnetic moment, and radii. Although the massive constituent quarks do not provide a complete picture, the inclusion of the meson cloud, which corresponds to Goldstone bosons from spontaneous chiral symmetry breaking in Quantum Chromodynamics (QCD), makes the constituent quark model a acceptable approach. Models such as the cloudy bag model \cite{Lu:1997sd, PhysRevC.66.032201:Miller} and
the chiral quark models \cite{OSET1984456, PhysRevD.73.114021:chiral1} have been applied successfully to study nucleon form factors.

In parallel, the dispersion theory has been developed to extract electromagnetic form factors of baryons. This approach is particularly effective at low energies ($<1\text{ GeV}$) where the  intermediate states can be conveniently isolated. Significant applications of the dispersion theory began with the inclusion of uncorrelated two-pion spectral function in studying nucleon form factors  \cite{PhysRevLett.2.365_FF1,PhysRev.117.1609_FF2,PhysRev.117.1603_FF3}. For overview and more insights, we may refer to Ref. \cite{Lin_2021:review_dispersion}. 
Regarding the proton radius puzzle, as being reported by the Particle Data Group (PDG) \cite{ParticleDataGroup:2022pth}, the smaller charge radius is more widely accepted because of  the precision of three measurements \cite{Antognini:2013txn:smallrp1,Xiong:2019umf:smallrp2,Bezginov:2019mdi:smallrp3}. The results from dispersion relation analyses \cite{Belushkin:2006qa:rpdisper,Hoferichter_2016_pipi_con, Lin:2021xrc_new_insight} are in good agreement with the experimental data. 

In this work, we take advantage of the dispersion relation to extract, from experimental data, the meson cloud contribution to nucleon form factors, and calculate the contribution of the quark core in a quark model. The nucleon structure in the core and cloud sectors may be extracted by fitting the combined contribution of the meson cloud and quark core to experimental data.
The paper is arranged as follows: The dispersion relation formalism of nucleon isospin form factors is given in Section \ref{sec:2}. The meson cloud contribution to nucleon vector form factors is calculated in Section \ref{sec:3}. Results from combining the meson cloud and quark core contributions are given in Section \ref{sec:4}. Conclusions are in Section \ref{sec:5sum}.

\section{\label{sec:2} Form Factors in Dispersion Relation Framework}

In the isospin symmetry limit, the isovector form factors of nucleon are defined as  
\begin{align}\label{eq:defFF}
  \langle 0 \vert j^\mu \vert (p \bar p - n \bar n)/2 \rangle &= \\ e \, \bar v_N \, &\left( \gamma^\mu \, F_1(q^2) - \frac{i \sigma^{\mu\nu} \, q_\nu}{2 m_N} \, F_2(q^2) 
  \right) \, u_N \nonumber
\end{align}
with
\begin{eqnarray}
  G_E(q^2) & = & F_1(q^2) + \frac{q^2}{4 m_N^2} \, F_2(q^2) \,, 
  \nonumber \\ 
  G_M(q^2) & = & F_1(q^2) + F_2(q^2)  \,.
  \label{eq:defFFEM}
\end{eqnarray}
where $q^2$ denotes the square of the invariant mass of the virtual photon, $G_{M}$ and $G_{E}$ are the magnetic and electric vector form factors, respectively. With the definition in (\ref{eq:defFF}), one may theoretically obtain $G_E(0)=\frac12$, $G_M(0)=\frac12(1+\kappa_p-\kappa_n)\approx 2.35$, where $\kappa_{p/n}$ denotes the magnetic moment of the proton/neutron. 

\begin{figure}
    \includegraphics[keepaspectratio,width=0.6\columnwidth]{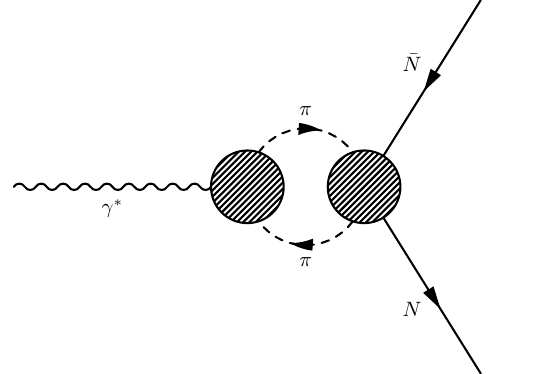}
    \caption{Two-pion continuum spectral of the isovector scattering channel.}
    \label{fig:twopioncon}
\end{figure}

By applying the Cauchy integral formula and the Schwarz reflection principle, one may link the entitle form factors to their imaginary parts, 
\begin{equation}
	G_{M/E}(s) = 
	\frac{1}{\pi} \, \int\limits_{s_0}^\infty ds^\prime \, 
	\frac{\operatorname{Im} G_{M/E}(s^\prime)}{s^\prime-s}  \,.
	\label{eq:disperunsubtr} 
\end{equation} 
In principle, all possible intermediate states between the virtual photon and the $\overline NN$ states should be included to calculate the $\operatorname{Im} G_{M/E}$ in the approach of dispersion relation. In the present case, however, the only important intermediate state is two-pion continuum states as depicted in Fig. \ref{fig:twopioncon} since we are handling a low energy problem. 
Therefore, we may set the lower bound in (\ref{eq:disperunsubtr}) to be $s_0=4m_\pi^2$, and the imaginary part of the form factors are derived \cite{Hoferichter_2016_pipi_con}, 
\begin{eqnarray}
    \operatorname{Im} G_{M}(s) &=& \frac{p_{\rm c.m.}^3(s)}{\sqrt{2s}}\, F^{V*}_\pi(s) \, f^1_{-}(s) \,, \nonumber \\
    \operatorname{Im} G_{E}(s) &=& \frac{p_{\rm c.m.}^3(s)}{m_N\sqrt{s}}\, F^{V*}_\pi(s) \, f^1_{+}(s) \,,
    \label{eq:ImGME}
\end{eqnarray}
where $p_{\rm c.m.}$ denotes the pion momentum in the center-of-mass frame of the two-pion system, $f^1_\pm$ are $\pi\pi\rightarrow N\Bar{N}$ partial wave transition amplitudes with the total angular momentum $J=1$ and the subscript $\pm$ referring to parallel/antiparallel antinucleon-nucleon helicity, and the pion vector form factor, $F^{V}_\pi$, is defined in the equation 
\begin{equation}
  \label{eq:pionFFdef}
  \langle 0 \vert j^\mu \vert \pi^+(p_+) \, \pi^-(p_-) \rangle = e \, (p_+^\mu -p_-^\mu) \, F^V_\pi((p_++p_-)^2)  \,. \phantom{m}
\end{equation}

The pion vector form factor $F^V_\pi$ as defined in (\ref{eq:pionFFdef}) may be determined as given in \cite{Hoferichter_2016_RS_ana}, and can be approximated as \cite{Leupold_2018}
\begin{eqnarray}
  \label{eq:FV-Omnes-alphaV}
  F^V_\pi(s) = (1+\alpha_V \, s) \, \Omega(s) \,,
\end{eqnarray}
where the Omn\`es function, $\Omega(s)$ is defined as
\begin{eqnarray}
  \Omega(s) = \exp\left\{ s \, \int\limits_{4m_\pi^2}^\infty \frac{ds'}{\pi} \, \frac{\delta(s')}{s' \, (s'-s-i \epsilon)} \right\} \,.
  \label{eq:omnesele}  
\end{eqnarray}
To ensure that $\operatorname{Im} G_{M/E}$ stays real, the phase shift used for the pion form factor calculation must coincide with the $\pi\pi\rightarrow N\Bar{N}$ partial wave, according to Watson's theorem \cite{watson_Theorem}. 
\begin{figure}
    \centering
    \begin{subfigure}[t]{\columnwidth}    \includegraphics[keepaspectratio,width=0.8\columnwidth]{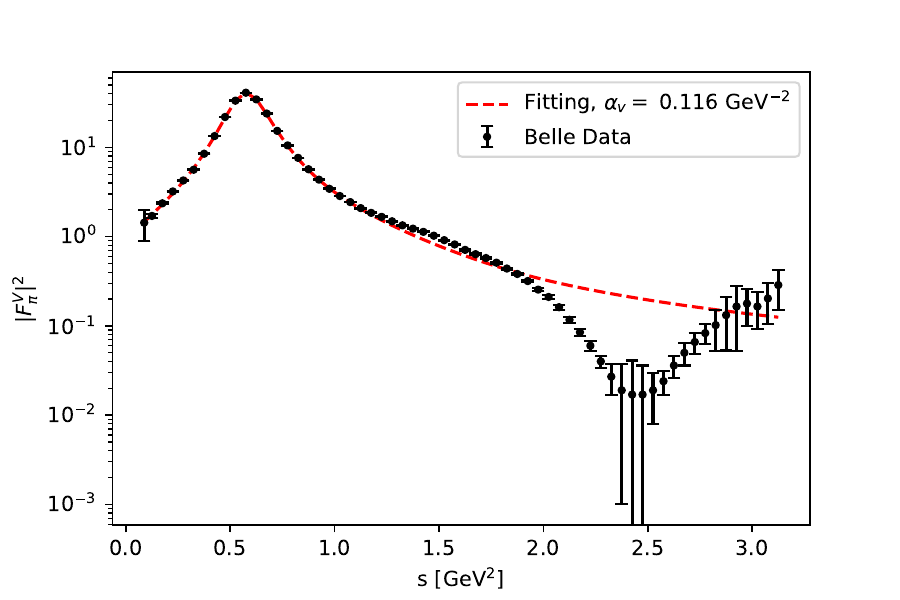}
    \caption{Modulus squared}
    \end{subfigure}
    \begin{subfigure}[t]{\columnwidth}
    \includegraphics[keepaspectratio,width=0.8\columnwidth]{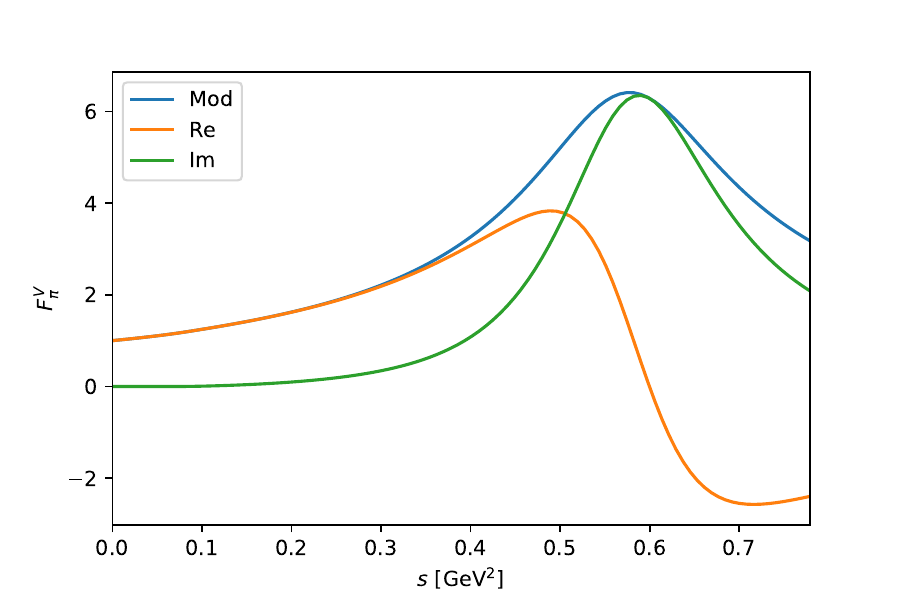}
    \caption{All components}
    \end{subfigure}
    \caption{Fitted pion vector form factor (modulus squared) with data obtained from Belle Experiment\cite{Fujikawa_2008}, and all components of the form factor.}
    \label{fig:pionVFF}
\end{figure}
By applying the phase shift $\delta(s)$ given by \cite{Garcia-Martin:2011iqs} and fitting $F^V_\pi(s)$ in (\ref{eq:FV-Omnes-alphaV}) to the data in Fig. \ref{fig:pionVFF}a, one gets  
\begin{eqnarray}
  \label{eq:alphaV}
  \alpha_V = 0.116 \, {\rm GeV}^{-2} \,.
\end{eqnarray}
The full pion vector form factor resulting from eqs. (\ref{eq:FV-Omnes-alphaV}) and  (\ref{eq:alphaV}) is shown in Fig. \ref{fig:pionVFF}b.

\section{\label{sec:3} meson cloud contribution from $\pi\pi \rightarrow N\Bar{N}$ transition amplitudes}

\begin{figure}
	\centering
	\begin{subfigure}[t]{0.40\columnwidth}
		\includegraphics[keepaspectratio,width=\columnwidth]{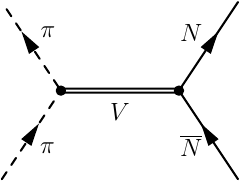}
		\caption{$\Tilde{f}^1_\pm$}
	\end{subfigure}
	\qquad
	\begin{subfigure}[t]{0.32\columnwidth}
		\includegraphics[keepaspectratio,width=\columnwidth]{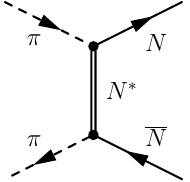}
		\caption{$f^1_{\pm N}$}
	\end{subfigure}
	\caption{Scattering processes correspond to $\Tilde{f}^1_\pm$ and $f^1_{\pm N}$ partial wave amplitudes.}
	\label{fig:pw_components}
\end{figure}

In this section, we extract the meson cloud contribution from $\pi\pi \rightarrow N\Bar{N}$ partial wave transition amplitudes, based on the data in Ref. \cite{LandoltBornstein1983:data_book}. The data we brought to consider here are $f^1_\pm,\, \Tilde{f}^1_\pm,\,\text{ and } f^1_{\pm N}$.
As shown in Fig. \ref{fig:pw_components}, $f^1_{\pm N}$ is the projection of nucleon pole terms, which contains only real part, and $\Tilde{f}^1_\pm$ results from the vector dominant process. $f^1_{\pm N}$ and $\Tilde{f}^1_\pm$ together give the total partial wave transition amplitude, $f^1_\pm$, 
\begin{equation}
	f^1_\pm = \Tilde{f}^1_\pm + f^1_{\pm N} \,
\end{equation} 
In this work, we assume that the nucleon pole terms regard to the meson cloud contribution at the low energy or low momentum transfer. Considering that more precise calculations of the total $\pi\pi \to N\Bar{N}$ partial wave amplitudes have been performed in \cite{Hoferichter_2016_RS_ana} in the framework of Roy–Steiner (RS) equations, we rather employ the RS solutions for $\operatorname{Im} G_{M/E}$. Accordingly, we scale the nucleon pole terms $f^1_{\pm N}$ with the $\operatorname{Im} G_{M/E}$ results in Ref. \cite{Hoferichter_2016_RS_ana}.

\begin{figure}
\includegraphics[keepaspectratio,width=\columnwidth]{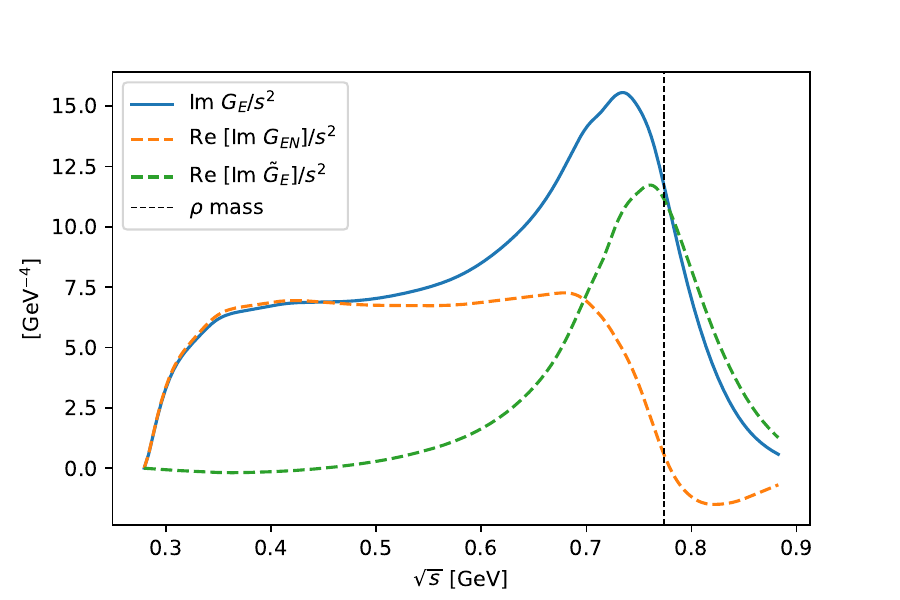}
\caption{Isovector spectral function together with its components divided by $s^2$.}
\label{fig:SpectralLines}
\end{figure}

It is found that (\ref{eq:ImGME}) does not lead to a real $G_E$ or $G_M$ by applying the full pion vector form factor, $F_\pi^V$, derived in the previous section and the $f_{\pm N}^1$ or $\Tilde{f}^1_\pm$ alone. The reason is the uncorrelation between the pion form factor and the partial waves $f_{\pm N}^1$ or $\Tilde{f}^1_\pm$. The pion form factor corrected to the $f_{\pm N}^1$ should be the one without $\rho$ meson contribution. 
Detailed calculations show that the imaginary parts of $\operatorname{Im} G_{M/E}$, derived in (\ref{eq:ImGME}) when applying $f_{\pm N}^1$ and $\Tilde{f}^1_\pm$ respectively,  cancel each other. That is, 
\begin{equation}
    \operatorname{Im} G_{M/E} = \operatorname{Re} [\operatorname{Im} \Tilde{G}_{M/E}] + \operatorname{Re} [\operatorname{Im} G_{M/E\,N}] \,,
\end{equation}
where $\operatorname{Im} \Tilde{G}_{M/E}$ and $\operatorname{Im} G_{M/E\,N}$ are extracted from (\ref{eq:ImGME}) when using only $\Tilde{f}^1_\pm$ and $f_{\pm N}^1$, respectively.
Therefore, these real parts from two different processes represent meaningful contributions of nucleon electromagnetic form factors, regardless of their imaginary parts.
The isovector spectral function and its components, divided by $s^2$, are shown in Fig. \ref{fig:SpectralLines}. It clearly shows that $\operatorname{Im} \Tilde{G}_{E}$ is the contribution from $\rho$ meson, leaving $\operatorname{Im} G_{M/E\,N}$ the contribution of uncorrelated two pions. 

\begin{figure}
    \centering
        \begin{subfigure}[t]{0.48\columnwidth}
        \includegraphics[keepaspectratio, width=\columnwidth]{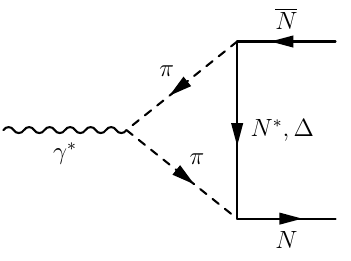}
        \caption{Time-like Region}
    \end{subfigure}
    \quad
    \begin{subfigure}[t]{0.45\columnwidth}
        \includegraphics[keepaspectratio, width=\columnwidth]{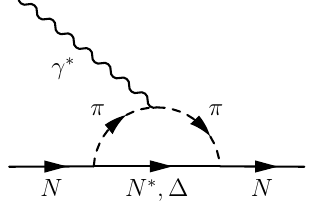}
	\caption{Space-like Region}
    \end{subfigure}
    \caption{Scattering processes correspond to the electromagnetic form factors determined at time-like region ($q^2>4m_N^2$) and space-like region ($q^2<0$).}
    \label{fig:feynman}
\end{figure}

\begin{figure}
\includegraphics[keepaspectratio,width=\columnwidth]{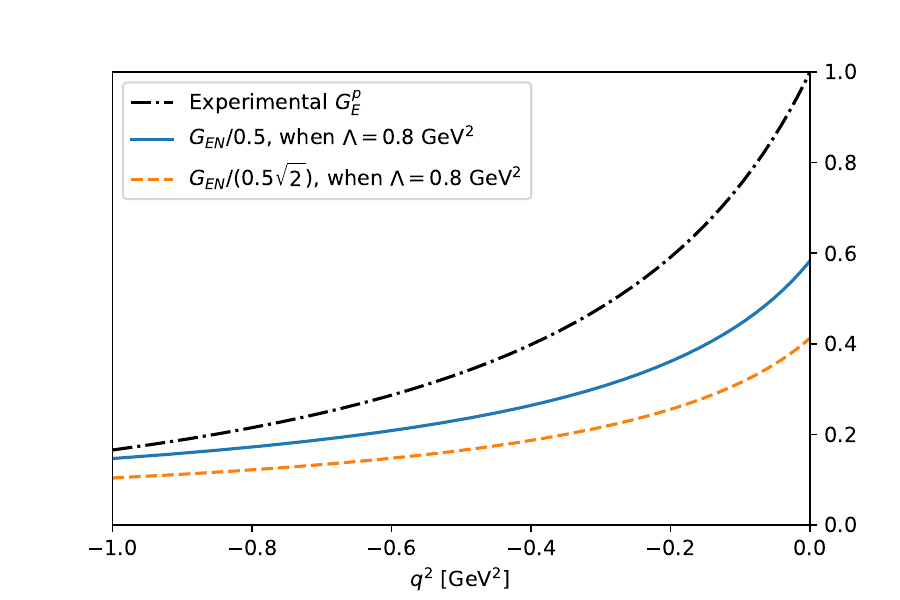}
\caption{Meson cloud contribution (electric form factor at space-like region) compared to the scale of dipole form of electromagnetic form factor of nucleon.}
\label{fig:GEdipolecom}
\end{figure}

The analyticity of dispersion relation formalism allows extracting the space-like values from the time-like ones. In this work, we extract the space-like $G_{M/E}$ from data in the time-like process, $\gamma^*\rightarrow N\overline N$, as illustrated in Fig. \ref{fig:feynman}. Inserting into (\ref{eq:disperunsubtr}) the $\operatorname{Re} [\operatorname{Im} G_{M/E\,N}]$ in Fig. \ref{fig:SpectralLines}, we derive the meson-cloud contribution to the electric vector form factors in the space-like region, as shown in Fig. \ref{fig:GEdipolecom}. Note that we have multiplied the $G_{EN}$ by a factor 2 to present it in the scale of proton electric form factor. In the calculation, we have applied an integral cut-off $\Lambda=0.8\,\rm GeV^2$ to the integration in (\ref{eq:disperunsubtr}) as it is the upper limit of experimental data in the market at present. In Fig. \ref{fig:GEdipolecom}, the solid curve shows the result directly derived from (\ref{eq:ImGME}) while the dashed curve shows the result which is divided by an extra factor $\sqrt{2}$ \footnote{\label{fn:factor} We are afraid that the formula in (\ref{eq:ImGME}) may not be consistent with the definition in (\ref{eq:defFF}). It is expected that the transition amplitudes, $f^1_\pm$ is defined for the initial state
\begin{eqnarray}
|\pi\pi,I=1,I_z=0\rangle	
\end{eqnarray}  
and the final state
\begin{eqnarray}
	|N\overline N, I=1,I_z=0\rangle	= (p \bar p - n \bar n)/\sqrt{2}
\end{eqnarray} 
Therefore, the formula in (\ref{eq:ImGME}) might result in a value larger than the definition by a factor of $\sqrt{2}$.}. 

The ideal value of the nucleon isovector electric charge is given by
\begin{eqnarray}
    G_E(0)&=&\frac{G^p_E(0)-G^n_E(0)}{2} \nonumber \\
    &=& \frac{Q^p - Q^n}{2} = 0.5 \,,
\end{eqnarray}
where $Q^p$ and $Q^n$ are charge number of proton and neutron, respectively. Our calculation leads to $G_E(0) = 0.766$, which is in a good agreement with the results in \cite{Hoferichter_2016_pipi_con}. There are ongoing discussions on correcting this number, such as modifying the integral cut-off \cite{Hoferichter_2016_pipi_con} or including inputs from chiral perturbation theory \cite{Leupold_2018}. One alternative discussion is also included in the footnote\footref{fn:factor}.

\section{\label{sec:4} Form factors from quark core and meson cloud contribution}

In the quark model regime, the proton can be considered as a core of three constituent quarks ($q^3$). Including the meson cloud (MC), the electric form factor of proton may be written as
\begin{equation}
    G_E = G_{E}^{q^3} + G_E^{\text{MC}}\,,
\end{equation}
with
\begin{alignat}{3}
    G_E^{\text{MC}}&=\
    && \frac{G_{EN}}{0.5} \,\\
    G_{E}^{q^3}&=\ &&\Braket{N',S'_z=1/2|q_1'q_2'q_3'}\bra{q_1'q_2'q_3'}\sum_{i=1}^3\mathcal{\hat I}_i\ket{q_1q_2q_3}\nonumber\\
    &&&\Braket{q_1q_2q_3|N,S_z=1/2},
\end{alignat}
where \(\ket{N, S_z}\) and \(\ket{N', S'_z}\) represent the initial and final states of the nucleons with their respective spin projections $S_z$ and $S'_z$. The operator $\mathcal{\hat I}_i$ describes the interaction of the photon with the quark $i$ in electromagnetic transition $\gamma q_i \to q_i'$. The completeness relations \(\ket{q_1' q_2' q_3'} \bra{q_1' q_2' q_3'}\) and \(\ket{q_1 q_2 q_3} \bra{q_1 q_2 q_3}\) are inserted to project all possible degrees of freedom of the initial and final nucleon states in the three-quark picture. This means that one needs to sum over spin, flavor, and color, and integrate over all momenta of the quarks. The spin, flavor, and color parts of the nucleon wave function are characterized by spin \(SU(2)\), flavor \(SU(2)\), and color \(SU(3)\) symmetries, respectively. For the spatial part, we expand it in the harmonic oscillator basis and extract the quark distribution of the quark core by comparing to the proton electric form factor. For more details, we refer the reader to our previous works about the study of \(N(1440)\), \(N(1520)\), and \(N(1535)\) \cite{Atthaphon:2021, Kaewsnod2022}.

\begin{figure}[t]
\includegraphics[keepaspectratio,width=\columnwidth,trim={0 0.5cm 0 1.5cm},clip]{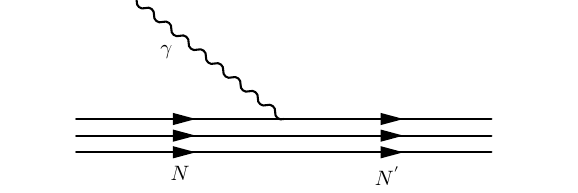}
\caption{The quark line diagram for the quark core contribution in $N\gamma\to N$.}
\label{fig:quarkcore}
\end{figure}

The electromagnetic transition of the quark core contribution is approximated in the impulse approximation, as shown in Fig. \ref{fig:quarkcore}, where the interaction of the photon with quark \(i\) is described by the coupling of the quark current to the electromagnetic field in the Breit frame as follows:
\begin{alignat}{2}
    &\bra{q_1'q_2'q_3'}\mathcal{\hat I}_i\ket{q_1q_2q_3} \nonumber \\
    =\ &\bra{q'_i}Q_i \bar{u}_{s'}(p')\gamma^\mu u_s(p)\epsilon_\mu(k)\ket{q_i} \Braket{q'_jq'_k|q_jq_k}.
\end{alignat}
Here, $i=1, 2, 3$ represents one of the quarks, while $j$ and $k$ denote the other two quarks, \(Q_i\) represents the electric charge of the quark \(i\), and the polarization vector of the photon is defined as \(\epsilon_\mu = (1, 0, 0, 0)\).

\begin{figure*}
    \centering
    \begin{subfigure}[t]{0.8\columnwidth}
    \includegraphics[keepaspectratio,width=\columnwidth]{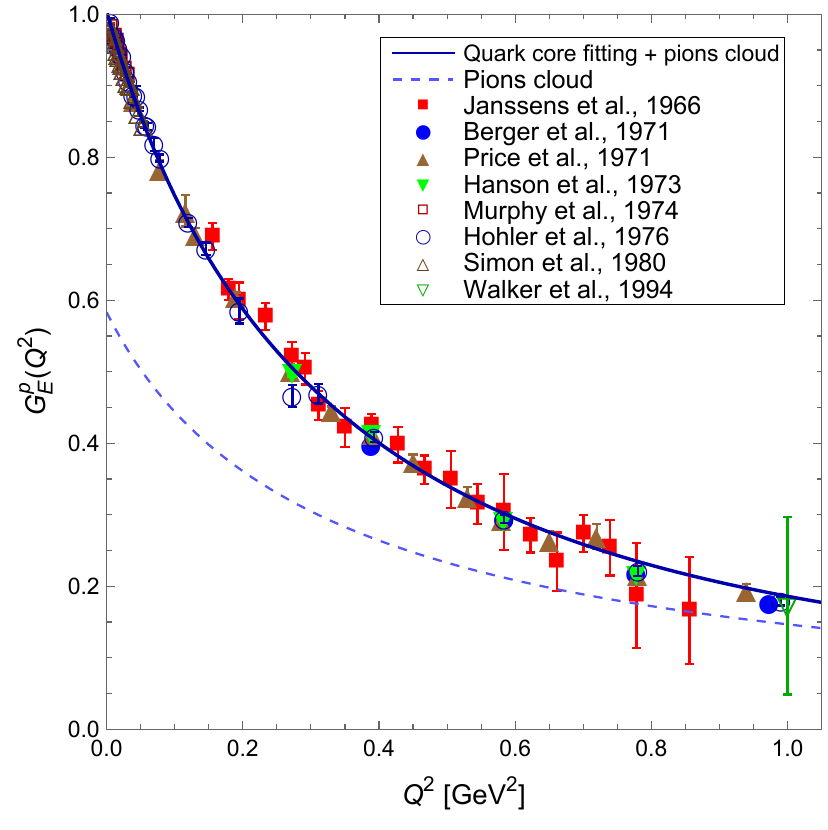}
    \caption{$\Lambda=0.8\text{ GeV}^2$}
    \end{subfigure}
    \qquad
    \begin{subfigure}[t]{0.8\columnwidth}	\includegraphics[keepaspectratio,width=\columnwidth]{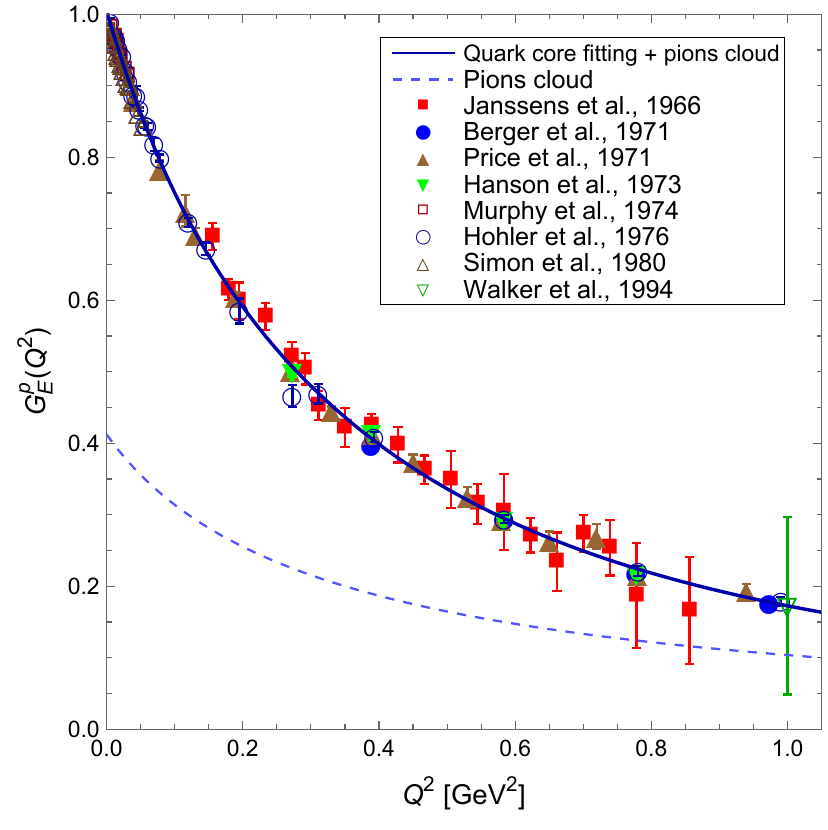}
	\caption{$\Lambda=0.8\text{ GeV}^2$ and $G_{EN}$ scaled by $1/\sqrt{2}$}
    \end{subfigure}
\caption{Proton electric form factors given by combining meson cloud contribution from dispersion relation and quark core contribution from quark model fitting, compared to experimental data \cite{PhysRev.142.922:pFF1,BERGER197187:pFF2,PhysRevD.4.45:pFF3,PhysRevD.8.753:pFF4,PhysRevC.9.2125:pFF5,HOHLER1976505:pFF6,SIMON1980381:pFF7,PhysRevD.49.5671:pFF8}.}
\label{fig:GEp}
\end{figure*}
\begin{figure}
    \centering
    \includegraphics[width=0.8\columnwidth]{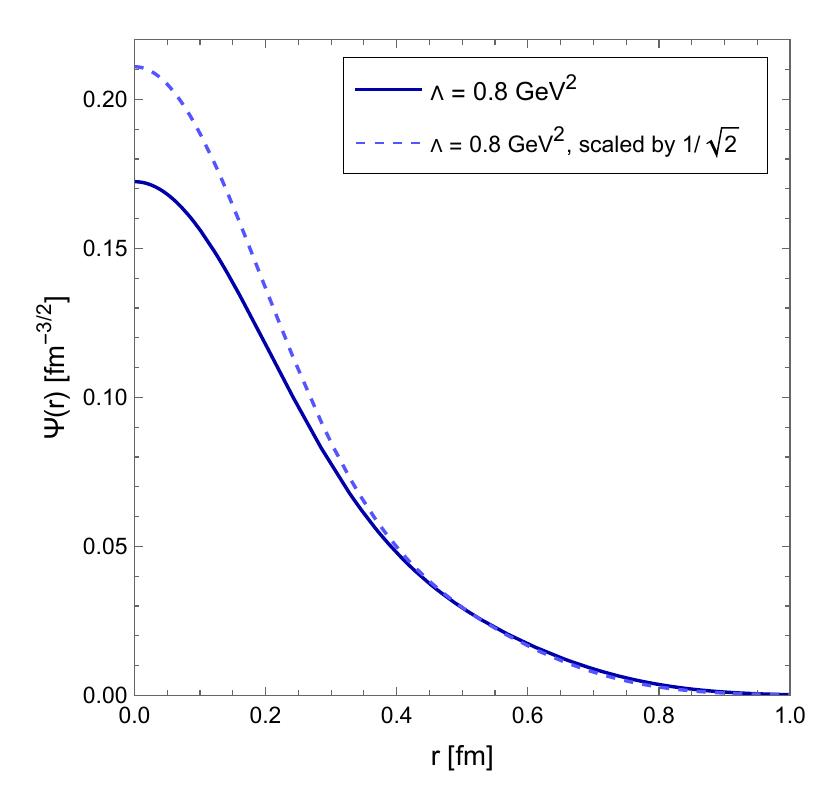}
    \caption{Spatial wave functions of quark core as a function of relative distance between the quark to the center of mass.}
    \label{fig:wave_function}
\end{figure}
The spatial wave function of the quark core is fitted to experimental data of the proton electric form factor, $G_E^p$, considering for two cases of the meson cloud contribution.   Fig. \ref{fig:GEp}a and \ref{fig:GEp}b show the fitting results with the meson cloud contributions $G_{EN}$ as the solid and dashed curves in Fig. \ref{fig:GEdipolecom}, respectively. The corresponding wave functions are shown as the solid and dashed curves in Fig. \ref{fig:wave_function}, respectively. The square charge radius of the proton and its quark core is calculated in the formulae,
\begin{eqnarray}
    \langle r^2_E \rangle^p &=& -6\Bigg[ \frac{d}{dq^2}G_E(q^2) \Bigg]_{q^2=0} \nonumber \\
    \langle r^2_E \rangle^{p,q^3} &=& -6\Bigg[ \frac{d}{dq^2}G_{E}^{q^3}(q^2) \Bigg]_{q^2=0}
\end{eqnarray}
The results of the mean square charge radii are given in Table \ref{tab:radii}.
It is found that the theoretical charge radius of the proton, with the $G_{EN}$ scaled by the factor $1/\sqrt{2}$, agrees well with the usual results from $ep \to ep$ experiments, as it was fitted directly, regardless of the small charge radius given by PDG \cite{ParticleDataGroup:2022pth}, $r_E^p = 0.8409 \pm 0.0004 \text{ fm}$. 
\begin{table}[h]
\caption{\label{tab:radii}%
Charge radii from 
two conditions used to determine meson cloud contribution of electromagnetic form factors.
}
\begin{ruledtabular}
\begin{tabular}{lll}
\textrm{Conditions}&
\textrm{$r_E^{p}\text{ [fm]}$}&
\textrm{$r_E^{p,q^3}\text{ [fm]}$}\\
\colrule
$\Lambda=0.8\text{ GeV}^2$, $G_{EN}$ from (\ref{eq:ImGME})& 0.888 & 0.586 \\
$\Lambda=0.8\text{ GeV}^2$, $G_{EN}$ scaled by $1/\sqrt{2}$ & 0.874 & 0.669 \\
\end{tabular}
\end{ruledtabular}
\end{table}

\section{\label{sec:5sum}Summary}
In this study, we have explored the meson cloud contribution to the nucleon electromagnetic form factors in dispersion relation approach. The meson cloud contribution is extracted by analyzing experimental data on pion-nucleon scatterings and isolating the nucleon pole terms.
The proton electromagnetic form factors are well reproduced in the work by combining the meson cloud contributions from dispersion relation and quark core contributions from quark model.  

Although one can fit the experimental proton electric form factor with the meson cloud contribution derived in (\ref{eq:ImGME}), the meson cloud contribution appears too large and the resulting quark core is rather small.

\begin{acknowledgments}
This research has received funding support from the NSRF via the Program Management Unit for Human Resources \& Institutional Development, Research and Innovation [grant number B50G670107].
\end{acknowledgments}



\nocite{*}
\bibliography{References}

\end{document}